\documentclass[journal]{IEEEtran}
\usepackage{subfigure}
\usepackage{color}
\usepackage{amsmath}
\usepackage{cite}%
\usepackage{graphicx}
\usepackage{epstopdf}
\usepackage{amsfonts,amsmath,amssymb}
\usepackage{graphicx}
\usepackage{url}
\usepackage{epstopdf}
\usepackage{bm}
\usepackage{bbm}
\usepackage{subfigure}
\usepackage{stfloats}
\usepackage{amsthm,amsmath,amssymb}
\usepackage{mathrsfs}
\usepackage{algorithm}
\usepackage{algorithmic}

\begin{document}
\newtheorem{corollary}{\textbf{Corollary}}
\newtheorem{remark}{\textbf{Remark}}
\newcommand{\Hnull}{\mathcal{H}_0}
\newcommand{\Halt}{\mathcal{H}_1}
\newcommand{\Hini}{\mathcal{H}_i}
\newcommand{\Houtj}{\mathcal{D}_j}
\newcommand{\Honull}{\mathcal{{D}}_0}
\newcommand{\Hoalt}{\mathcal{{D}}_1}


\title{Hybrid Relay-Reflecting Intelligent Surface-Aided Covert Communications}
\author{{Jinsong Hu, Xiaoqiang Shi, Youjia Chen, Tiesong Zhao, and Feng Shu}


}

\maketitle

\begin{abstract}
In this paper, we propose to use a hybrid relay-reflecting intelligent surface (HR-RIS) to enhance the performance of a covert communication system. Specifically, the HR-RIS consists of passive reflecting elements and active relay elements to manipulate the wireless signals from a transmitter to a desired receiver while ensuring the covertness of the transmission via avoiding such signals being detected by a warden. To fully explore the benefits offered by the HR-RIS, we first formulate the joint design of the transmit power and relay/reflection coefficients of the HR-RIS as an optimization problem to maximize the covert rate subject to a covertness constraint. To tackle the solution to this optimization problem, we then derive a closed-form expression for an upper bound on covert rate, based on which we develop an alternate algorithm to solve the formulated optimization problem. Our examination shows that the HR-RIS outperforms the traditional RIS in term of achieving a higher covert rate. Interestingly, we also observe the major part of the performance gain brought by the HR-RIS can be obtained by a small number of active relay elements (e.g., 5) and further increasing this number does not improve the covert communication performance.
\end{abstract}

\section{Introduction}\label{sec:introduction}
Covert communication, or low probability of detection (LPD) communication, has emerged as a cutting-edge secure communication technique aiming at hiding communications from a watchful warden \cite{2019Low}. Considering different practical constraints and limitations, covert communication in different scenarios was widely studied in the past few years. For example, covert communication in relay networks was studied, and it was proved that the relay can stealthily transmit its own information to the desired destination in the process of forwarding data \cite{Hu2018covertrelay}.
In \cite{2020Intelligent}, the authors presented the potentials of using the recently emerged intelligent reflecting surface (IRS), also known as reflecting intelligent surface (RIS), to improve covert communication performance.

Following \cite{2020Intelligent}, several RIS-aided covert communication approaches have been developed.
Specifically, RIS is a flat surface composed of a large number of reconfigurable and low-cost passive reflective elements, each of which is capable of controlling the phase and amplitude of the incident signal for optimal reflection, making the wireless channel between the transmitter and receiver more favorable for communication \cite{2020bo}.
For example,
the authors of \cite{2021MIMO} investigated the multiple-input-multiple-output (MIMO) covert communication assisted by RIS, where the covert rate was maximized by jointly designing the transmit covariance matrix and phase shift matrix.
In \cite{2020DC}, the authors considered the design of a latency constrained covert communication system with the assistance of RIS with global CSI and without the warden's instantaneous CSI, respectively.

It should be emphasized that the existing studies on RIS-assisted covert communication adopted the fully-passive beamforming strategies. A main limitation of the traditional RIS compared to
relays is the that the passive reflection limits the degrees of freedom in the beamforming. Thus, the traditional RIS cannot outperform a half-duplex relay when the number of the elements in the RIS is not sufficiently large \cite{2019Intelligent}. This implies that if a few passive elements of the RIS are replaced by active ones, the traditional RIS becomes a hybrid relay reflecting intelligent surface (HR-RIS) \cite{2021Hybrid}, which may lead to that the HR-RIS is able to significantly improve the assisted system performance. Therefore, in this work we propose a HR-RIS-based covert communication scheme, which applies active beamforming by adding some relaying elements at the traditional RIS to mitigate the limitation of the passive reflecting.

The main contributions of this paper are summarized as follows. We consider the covert communication from a transmitter (Alice) to a receiver (Bob) with the aid of an HR-RIS. Our goal is to jointly optimize the transmit power and the HR-RIS reflection matrix, including its phase shifts and amplitudes, to maximize the covert rate at Bob. In the optimization problem of interest, the coefficient matrix of
HR-RIS is optimized by an alternate optimization (AO) method and then the optimal transmit power is obtained under the covertness constraint in terms of the Kullback-Leibler (KL) divergence. Our study shows that the HR-RIS can significantly improve the covert rate compared to conventional RIS.
\section{System Model}\label{sec:system_model}
\subsection{Considered Scenario and Adopted Assumptions}
As shown in Fig.~\ref{sYSm}, we propose a covert communication transmission scheme assisted by an HR-RIS, where a transmitter (Alice) intends to send confidential information to a legitimate receiver (Bob) with the aid of the HR-RIS, while a warden (Willie) attempts to detect the existence of this transmission. Specifically, there are two links from Alice to Bob, i.e., the direct link from Alice to Bob, and the reflection/relaying link form Alice to the HR-RIS and then to Bob. Similarly, there are two paths from Alice to Willie. Alice, Bob and Willie are assumed to be equipped with $N_{a}$, $N_{b}$, and $N_{w}$ antennas, respectively. In addition, it is assumed that the signals reflected by the HR-RIS twice or more are ignored due to the significant path loss \cite{2019irse}.

The HR-RIS is assumed to be equipped with $N$ elements, including $M$ passive reflecting elements and $K$ active relaying elements (i.e., $M + K = N$). The passive reflecting elements are implemented by phase shifter, while the active relaying elements can tune the phase and amplitude of the incident signals. We assume that the active elements work in the AF mode. Therefore, for $K=0$, HR-RIS converges to a traditional RIS. For $K=N$, by contrast, it becomes an AF relay station equipped with $N$ antennas. Hence, in this work we have $1\leq K \leq N$. Furthermore, similar to the conventional RIS, we assume that each (active/passive) element of HR-RIS can independently reflect the received signals.

For the HR-RIS, $\mathbb{Q}$  represents the set of active relay elements. We define $\mathbf{\Theta}=\mathbf{\Phi}+\mathbf{\Psi}$, where $\mathbf{\Theta}=\mathrm{diag}\{\theta_{1},\cdots,\theta_{N}\} \in \mathbb{C}^{N\times N}$ , $\mathbf{\Phi}=\mathrm{diag}\{\phi_{1},\cdots,\phi_{N}\} \in \mathbb{C}^{N\times N}$ , and {$\mathbf{\Psi}=\mathrm{diag}\{\psi_{1},\cdots,\psi_{N}\} \in \mathbb{C}^{N\times N}$ , where $\mathbf{\Phi}$ and $\mathbf{\Psi}$ denote the reflection coefficients of passive elements and active elements, respectively. Therefore, we have
\begin{align}
\mathbf{\theta}_{n}=
\left\{
  \begin{array}{ll}
     |\beta_{n}|e^{j\mu_{n}},  &\mathrm{if}\quad n\in \mathbb{Q}, \\
    e^{j\mu_{n}},&{\mathrm{otherwise,}}
  \end{array}
\right.
\end{align}
\vspace{-0.35cm}\begin{align}
\phi_{n}=
\left\{
  \begin{array}{ll}
     0,&\mathrm{if} \quad n\in \mathbb{Q}, \\
    e^{j\mu_{n}},& {\mathrm{otherwise,}}
  \end{array}
\right.
\end{align}
and
\vspace{-0.35cm}\begin{align}
\psi_{n}=
\left\{
  \begin{array}{ll}
     |\beta_{n}|e^{j\mu_{n}},&\mathrm{if} \quad n\in \mathbb{Q},\\
      0,& {\mathrm{otherwise,}}
  \end{array}
\right.
\end{align}
where $\mu_{n} \in [0,2\pi)$ represents the phase shift. We notice that $|\beta_{n}|=1$ for $n \notin \mathbb{Q}$, and $|\beta_{n}|$ for $n \in \mathbb{Q}$ is determined by the total power of the active elements, which will be discussed later.
\begin{figure}
  \centering
  \includegraphics[scale=0.35]{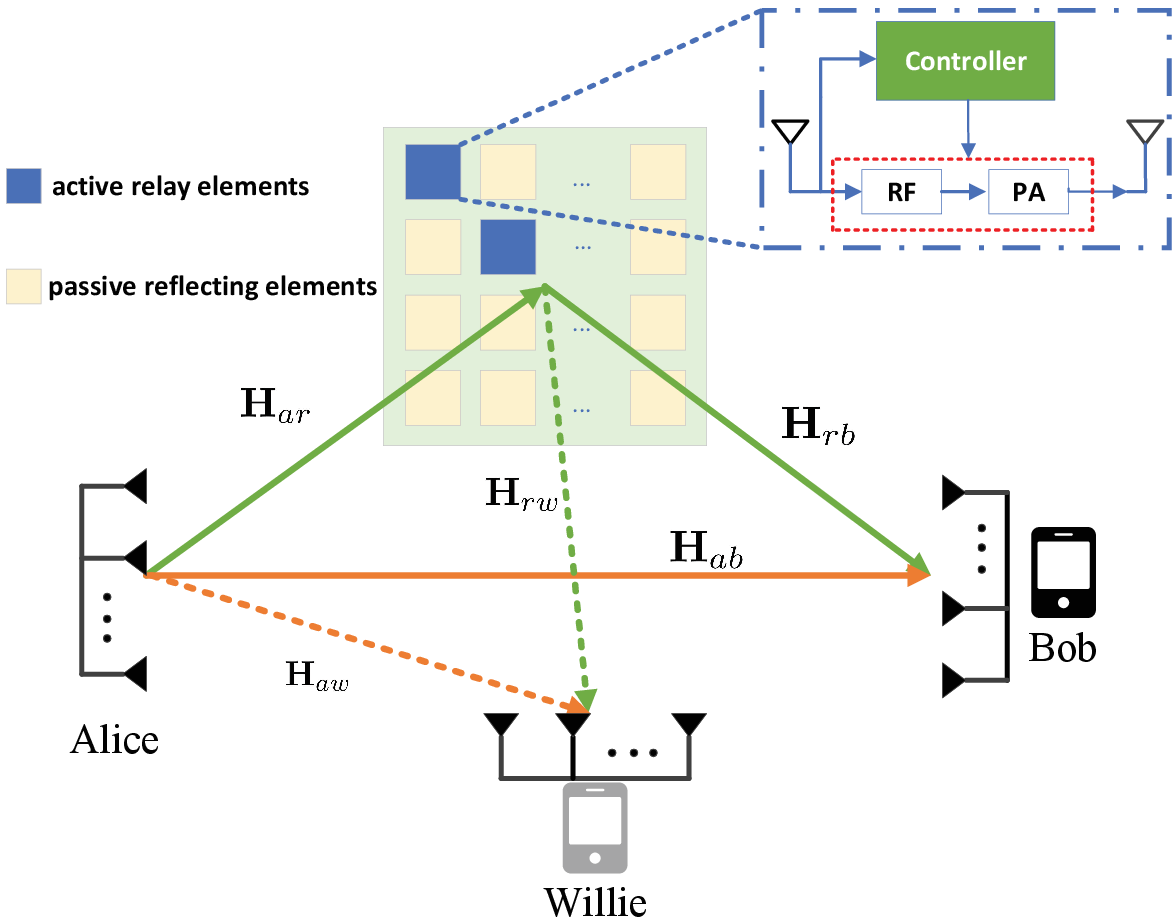}\\
  \caption{HR-RIS aided covert communication}\label{sYSm}
\end{figure}
\subsection{Transmission from Alice to Bob}
When Alice transmits confidential message, the received signal at Bob is given by
\begin{align}
\mathbf{y}_{b}= &\sqrt{P_{a}}\mathbf{H}_{rb}\mathbf{\Psi} \mathbf{H}_{ar}\mathbf{x}+\sqrt{P_{a}}\mathbf{H}_{rb}\mathbf{\Phi}\mathbf{H}_{ar}\mathbf{x} \notag \\
&+\sqrt{P_{a}}\mathbf{H}_{ab}\mathbf{x}+\mathbf{H}_{rb}\mathbf{\Psi} \mathbf{n}_{r}+\mathbf{n}_{b} \notag \\
&=\sqrt{P_{a}}(\mathbf{H}_{rb}\mathbf{\Theta}\mathbf{H}_{ar}+\mathbf{H}_{ab})\mathbf{x}
+\mathbf{n}_{bt},  \label{yb}
\end{align}
where $\mathbf{n}_{bt}=\mathbf{H}_{rb}\mathbf{\Psi} \mathbf{n}_{r}+\mathbf{n}_{b}$ represents the total effective noise at the Bob. $\mathbf{H}_{ar}$  and $\mathbf{H}_{ab}$ are the steering vectors from Alice to HR-RIS and the steering vector from Alice to Bob, respectively, and $\mathbf{H}_{rb}$ denotes the steering vector from HR-RIS to Bob. In addition, $\mathbf{n}_{r}$ $\sim$ $\mathcal{CN}(\mathbf{0},\sigma_{r}^{2}\mathbf{I}_{K})$ and $\mathbf{n}_{b}$ $\sim$ $\mathcal{CN}(\mathbf{0}, \sigma_{b}^{2}\mathbf{I}_{N_{b}})$ are the complex additive white Gaussian noise (AWGN) space vectors at the $K$ active elements of the HR-RIS and at the Bob, respectively. For simplicity, we assume that $\sigma_{r}^{2}=\sigma_{b}^{2}$ and $\mathbf{n}_{bt}$ $\sim$ $\mathcal{CN}(\mathbf{0},\sigma_{b}^{2}(\mathbf{I}_{N_{b}}+\mathbf{H}_{rb}\mathbf{\Psi} \mathbf{\Psi}^{H}\mathbf{H}_{rb}^{H}))$. $\mathbf{x}$ is the signal vector transmitted by Alice, with each elements following $\mathcal{CN}(0,1)$, and $P_{a}$ is Alice's transmit power.

Following \eqref{yb}, the transmit power of the active elements at the HR-RIS can be expressed as
\begin{align}
P_{r}=\mathrm{trace}(\mathbf{\Psi}(\mathbf{H}_{rb}\mathbf{H}_{rb}^{H}P_{a}
+\sigma_{b}^{2})\mathbf{\Psi}^{H}),
\end{align}
which should meet the constraint condition $P_{r}\leq P_{r}^{\mathrm{max}}$, where $P_{r}^{\mathrm{max}}$ is the maximum transmit power of all the $K$ active elements.
We assume that Bob's channel state information (CSI) is perfectly known by himself through channel estimation \cite{2020RISce}.
Following~\eqref{yb}, Bob's covert rate in the HR-RIS-aided covert communication system can be expressed as \cite{2021Hybrid}
\begin{align}
f(\mathbf{\Theta},P_{a})=\log_{2}|\mathbf{I}_{Nb}+\mathbf{\Omega}_{b}|, \label{problem}
\end{align}
where $\mathbf{\Omega}_{b}= {P_{a}\mathbf{U}_{b}\mathbf{R}^{-1}}/{\sigma_{b}^{2}}$, the signal covariance matrix is given by $\mathbf{U}_{b}=(\mathbf{H}_{rb}\mathbf{\Theta}\mathbf{H}_{ar}+\mathbf{H}_{ab})
(\mathbf{H}_{rb}\mathbf{\Theta}\mathbf{H}_{ar}+\mathbf{H}_{ab})^{H}$, and the aggregate noise covariance matrix is given by $\mathbf{R}=(\mathbf{I}_{N_{b}}+\mathbf{H}_{rb}\mathbf{\Psi} \mathbf{\Psi}^{H} \mathbf{H}_{rb}^{H})$ $\in$ $\mathbb{C}^{N_{b}\times N_{b}}$.
\subsection{Binary Hypothesis Testing at Willie}
In this work, we focus on the delay-constrained covert communication, that is, the number of channel uses $L$ is finite. In order to detect the existence of a transmission, Willie attempts to distinguish the following two hypotheses:
\begin{align}
\mathbf{y}_{w}=
\left\{
  \begin{array}{ll}
     \mathbf{n}_{w}, &\hbox{$\mathcal{H}_0$,} \\
    \sqrt{P_{a}}(\mathbf{H}_{rw}\mathbf{\Theta}\mathbf{H}_{ar}+\mathbf{H}_{aw})\mathbf{x}+\\
    \mathbf{H}_{rw}\mathbf{\Psi} \mathbf{n}_{r}+\mathbf{n}_{w}, & \hbox{$\mathcal{H}_1$,}    \label{yw}
  \end{array}
\right.
\end{align}
where $\mathbf{H}_{aw}$  and $\mathbf{H}_{rw}$ are the steering vector from Alice to Willie and the steering vector from HR-RIS to Willie, respectively, and $\mathbf{n}_{w}$ $\sim$ $\mathcal{CN}(\mathbf{0}, \sigma_{w}^{2}\mathbf{I}_{N_{w}})$ is the AWGN space vector at Willie. $\mathcal{H}_0$ denotes the null hypothesis in which Alice does not transmit, and $\mathcal{H}_1$ denotes the alternative hypothesis where Alice transmits signals. Similarly, we assume that $\sigma_{w}^{2}=\sigma_{r}^{2}$. Therefore, the total noise power at Willie is $\mathbf{n}_{wt}$ = ($\mathbf{H}_{rw}\mathbf{\Psi} \mathbf{n}_{r}+\mathbf{n}_{w}$) $\sim$ $\mathcal{CN}(\mathbf{0},\sigma_{w}^{2}(\mathbf{I}_{N_{w}}+\mathbf{H}_{rw}\mathbf{\Psi} \mathbf{\Psi}^{H}\mathbf{H}_{rw}^{H}))$ under $\mathcal{H}_1$.

Considering the worst-case scenario for covert communications, we assume that Willie knows the channels perfectly. The probability density function of $\mathbf{y}_{w}$ under $\mathcal{H}_0$ and $\mathcal{H}_1$ can be respectively given by
\begin{align}
f(\mathbf{y}_{w}|\mathcal{H}_0)=&\mathcal{CN}(\mathbf{0},\sigma^{2}_{w}\mathbf{I}_{N_{w}}),      \\
f(\mathbf{y}_{w}|\mathcal{H}_1)=&\mathcal{CN}(\mathbf{0},P_{a}\mathbf{U}_{w} +\sigma^{2}_{w}(\mathbf{I}_{N_{w}}+\mathbf{H}_{rw}\mathbf{\Psi} \mathbf{\Psi}^{H} \mathbf{H}_{rw}^{H})). \notag
\end{align}

As such, the signal-to-interference-plus-noise ratio (SINR) at Willie under $\mathcal{H}_1$ is given by
\begin{align}
\gamma_{w}=\frac{|\mathbf{U}_{w}|P_{a}}{(|\mathbf{M}|\sigma_{w}^{2})},
\end{align}where $\mathbf{U}_{w}=(\mathbf{H}_{rw}\mathbf{\Theta}\mathbf{H}_{ar}+\mathbf{H}_{aw})
(\mathbf{H}_{rw}\mathbf{\Theta}\mathbf{H}_{ar}+\mathbf{H}_{aw})^{H}$ and $\mathbf{M}=(\mathbf{I}_{N_{w}}+\mathbf{H}_{rw}\mathbf{\Psi}\mathbf{\Psi}^{H}\mathbf{H}_{rw}^{H})$. We note that $\mathbf{U}_{w}$ and $\mathbf{M}$ are Hermitian matrices.
Then, we perform the eigenvalue decomposition (EVD) on the above two matrices, which can be written as $\mathbf{U}_{w}=\mathbf{G}\mathbf{\Xi}\mathbf{G}^{-1}$ and $\mathbf{M}=\mathbf{J}\mathbf{\Lambda}\mathbf{J}^{-1}$, where $\mathbf{G}$ and $\mathbf{J}$ are matrices of eigenvectors of $\mathbf{U}_{w}$ and $\mathbf{M}$, respectively, and $\mathbf{G}$ and $\mathbf{J}$ $\in \mathbb{C}^{N_{w}\times N_{w}}$, $\mathbf{\Xi}=\mathrm{diag}\left\{\omega_{1}, \omega_{2},\cdots, \omega_{N_{w}}\right\}$, $\omega_{n}$  is the $n$-th eigenvalue of $\mathbf{U}_{w}$, $\mathbf{\Lambda}=\mathrm{diag}\left\{\kappa_{1},\kappa_{2},\cdots,\kappa_{N_{w}}\right\}$, $\kappa_{n}$  is the $n$-th eigenvalue of $\mathbf{M}$. As such, we have $|\mathbf{U}_{w}|=\prod_{i=1}^{N_{w}}\omega_{i}$ and $|\mathbf{M}|=\prod_{i=1}^{N_{w}}\kappa_{i}$.

In covert communications, $\mathcal{D}_{01}\leq 2\epsilon^{2}$ is generally adopted as the covertness constraint, where $\epsilon$ is a small value to determine the required covertness level and the KL divergence $\mathcal{D}_{01}$ is given by \cite{2018Delay}
\begin{align}
&\mathcal{D}_{01}=L\left[\ln(1+\gamma_{w})-
\frac{\gamma_{w}}{1+\gamma_{w}}\right]. \label{d01}
\end{align}
\subsection{Problem Formulation}
In this part, we jointly design the transmit power at Alice and relay/reflection coefficients of the HR-RIS to maximize the covert rate at Bob subject to the covertness and other constraints, of which the optimization problem can be formulated as
\vspace{-0.35cm}\begin{subequations}\label{a}
\begin{align}
\mathrm{(P1)}:&\mathop{\mathrm{max}}\limits_{\mathbf{\Theta},P_{a}} f(\mathbf{\Theta},P_{a}), \\
&\mathbf{s.t.}~\mathcal{D}_{01}\leq 2\epsilon^{2},\\
            &~~~~~|\mathbf{\beta}_{n}|=1,~\mathrm{for}~~n \notin \mathbb{Q}, \\
            &~~~~~P_{a}\leq P_{a}^{\mathrm{max}}.
\end{align}
\end{subequations}
\section{COVERT COMMUNICATION DESIGN}
In this section, our goal is to maximize the covert rate at Bob by jointly designing $P_a$ and $\mathbf{\Theta}$. We propose an alternate algorithm to optimize $P_a$ and $\mathbf{\Theta}$. Specifically, we first optimize $\mathbf{\Theta}$ for a given $P_{a}$ and the object function is transformed into a form that is easy to handle. Then, we optimize $P_{a}$} for a given $\mathbf{\Theta}$.
\subsection{Optimizing $\mathbf{\Theta}$ for a Given $P_{a}$}
First, we randomly generate the coefficient of HR-RIS and use $\mathcal{D}_{01}=2\epsilon^{2}$ to get the feasible $P_{a }$.
The object function $f(\mathbf{\Theta},P_{a})$ is non-convex with respect to $\mathbf{\Theta}$. In addition, the feasible set of (P1) is non-convex due to the unit-modulus constraint (\ref{a}c). Therefore, (P1) is difficult to be tackled. Thus, we approximate the objective function $f({\mathbf{\Theta}},P_{a})$ by using its upper bound $f_{0}({\mathbf{\Theta}},P_{a})$, which can be written as
\begin{align}
&f(\mathbf{\Theta},P_{a})=\log_{2}\left|\mathbf{I}_{N_{b}}+\frac{P_{a}\mathbf{U}_{b}\mathbf{R}^{-1}}
{\sigma_{b}^{2}}\right|\notag\\
&=\log_{2}|\mathbf{R}+\rho\mathbf{U}_{b}|-\log_{2}|\mathbf{R}|\notag\\
& \overset{a} \leq \log_{2}|\mathbf{R}  +\rho\mathbf{U}_{b}|\notag\\
&=\log_{2}\Big{|}\mathbf{R}  +\rho(\mathbf{H}_{rb}\mathbf{\Theta}\mathbf{H}_{ar}+\mathbf{H}_{ab})(\mathbf{H}_{rb}\mathbf{\Theta}\mathbf{H}_{ar}+\mathbf{H}_{ab})^{H}\Big{|}\notag\\
&=f_{0}(\mathbf{\Theta},P_{a}),
\end{align}
where $\rho=P_{a}/\sigma_{b}^{2}$ and $a$ is achieved by set $\mathbb{Q}=\emptyset$. We note that this upper bound becomes tighter as $\log_{2}|\mathbf{R}|$ decreases.
  Considering the case where the active elements employed at the HR-RIS are few, we know that $\mathbf{\Psi}$ is sparse, which means $\log_{2}|\mathbf{R}|$ is small as per its definition given \eqref{problem}. Therefore, for a given $P_{a}$ the optimization problem (P1) can be rewritten as
\begin{align}
\mathrm{(P2)}:&\mathop{\mathrm{max}}\limits_{\mathbf{\Theta}} f_{0}(\mathbf{\mathbf{\Theta}},P_{a}),\notag\\
&~\mathbf{s.t.}~(\mathrm{\ref{a}b}),(\mathrm{\ref{a}c}),(\mathrm{\ref{a}d}).
\end{align}
Although $f_{0}(\mathbf{\Theta},P_{a})$ is still non-convex, the optimal reflection coefficient of HR-RIS can be obtained by  our proposed method detailed in the following subsection.
\subsection{Transformation of the Objective Function}
Generally, the proposed solution is a sequential procedure where in each iteration, a specific coefficient of HR-RIS is updated when the others are fixed. Specifically, we let $\mathbf{a}_{n}^{H} \in \mathbb{C}^{N_{b}\times 1}$ denote the $n$-th row of $\mathbf{H}_{ab}$, and $\mathbf{b}_{n} \in \mathbb{C}^{N_{b}\times 1}$ denote the $n$-th column of $\mathbf{H}_{rb}$, i.e., $\mathbf{H}_{ar}=[\mathbf{a}_{1},\mathbf{a}_{2},\cdots,\mathbf{a}_{N}]^{H}$, and $\mathbf{H}_{rb}=[\mathbf{b}_{1},\mathbf{b}_{2},\cdots,\mathbf{b}_{N}]$. Since $\mathbf{\Psi}$ and $\mathbf{\Theta}$ are diagonal matrices, we have $\mathbf{H}_{rb}\mathbf{\Theta}\mathbf{H}_{ar}=\sum_{n=1}^{N}\mathbf{\theta}_{n}\mathbf{b}_{n}
\mathbf{a}_{n}^{H}$ and $\mathbf{H}_{rb}\mathbf{\Psi}=\sum_{n\in \mathbb{Q}}\mathbf{\theta}_{n}\mathbf{b}_{n}$. Hence, we can rewrite $f_{0}({\mathbf{\Theta}},P_{a})$ as
\begin{small}
\begin{align}
f_{0}(\mathbf{\Theta},P_{a})&=\log_{2}\left|\mathbf{R}+\rho(\mathbf{H}_{rb}\mathbf{\Theta}\mathbf{H}_{ar}+
\mathbf{H}_{ab})
(\mathbf{H}_{rb}\mathbf{\Theta}\mathbf{H}_{ar}+\mathbf{H}_{ab})^{H}\right|\notag\\
&=\log_{2}\Big{|}\mathbf{I}_{N_{b}}+\sum_{i\in Q}\mathbf{\theta}_{i}\mathbf{b}_{i}\mathbf{\theta}_{i}^{*}\mathbf{b}_{i}^{H}+\rho\sum_{i=1}^{N}
|\theta_{i}|^{2}\mathbf{b}_{i}\mathbf{a}_{i}^{H}\mathbf{a}_{i}\mathbf{b}_{i}^{H}\notag\\
&+\rho \mathbf{H}_{ab}\mathbf{H}_{ab}^{H}+\rho\sum_{i=1}^{N}\sum_{j=1,j\neq i}^{N}\mathbf{\theta}_{i}\mathbf{\theta}_{j}^{*}\mathbf{b}_{i}\mathbf{a}_{i}^{H}\mathbf{a}_{j}
\mathbf{b}_{j}^{H} \notag\\
&+\rho\sum_{i=1}^{N}(\mathbf{H}_{ab}\mathbf{\theta}_{i}^{*}\mathbf{a}_{i}\mathbf{b}_{i}^{H}
+\mathbf{\theta}_{i}\mathbf{b}_{i}\mathbf{a}_{i}^{H}\mathbf{H}_{ab}^{H})\Big{|}.
\end{align}
\end{small}

\vspace{-0.35cm}Consequently, $f_{0}({\mathbf{\Theta}},P_{a})$ can be rewritten as
\begin{align}
f_{0}({\mathbf{\Theta}},P_{a})=\log_{2}\left|\mathbf{A}_{n}+|\theta_{n}|^{2}\mathbf{B}_{n}
+\mathbf{\theta}_{n}\mathbf{C}_{n}
+\mathbf{\theta}_{n}^{*}\mathbf{C}_{n}^{H}\right|.
\end{align}

For $n\in \mathbb{Q}$, $\mathbf{A}_{n}$, $\mathbf{B}_{n}$ and $\mathbf{C}_{n}$ can be respectively written as
\begin{small}
\begin{align}
&\mathbf{A}_{n}=\mathbf{I}_{N_{b}}+\sum_{i\in \mathbb{Q},i\neq n}^{N}\mathbf{\theta}_{i}\mathbf{b}_{i}\sum_{i\in \mathbb{Q},i\neq n}^{N}\mathbf{\theta}_{i}^{*}\mathbf{b}_{i}^{H}\notag\\
&+\rho\left(\sum_{i=1 ,i\neq n}^{N}\mathbf{\theta}_{i}\mathbf{b}_{i}\mathbf{a}_{i}^{H}+\mathbf{H}_{ab}\right)\left(\sum_{i=1 ,i\neq n}^{N}\mathbf{\theta}_{i}\mathbf{b}_{i}\mathbf{a}_{i}^{H}+\mathbf{H}_{ab}\right)^{H},  \notag\\
&\mathbf{B}_{n}=\rho \mathbf{b}_{n}\mathbf{a}_{n}^{H}\mathbf{a}_{n}\mathbf{b}_{n}^{H}+\mathbf{b}_{n}\mathbf{b}_{n}^{H}, \notag\\
&\mathbf{C}_{n}=\mathbf{b}_{n}\sum_{i\in \mathbb{Q},i\neq n}^{N}\mathbf{\theta}_{i}^{*}\mathbf{b}_{i}^{H}+\rho  \mathbf{b}_{n}\mathbf{a}_{n}^{H}\left(\mathbf{H}_{ab}^{H}+\sum_{i=1 ,i\neq n}^{N}\mathbf{a}_{i}\mathbf{b}_{i}^{H}\mathbf{\theta}_{i}^{*}\right).  \notag
\end{align}
\end{small}
For $n\notin \mathbb{Q}$, $\mathbf{A}_{n}$, $\mathbf{B}_{n}$ and $\mathbf{C}_{n}$ are defined as
\begin{small}
\begin{align}
&\mathbf{A}_{n}=\mathbf{I}_{N_{b}}+\rho\left(\sum_{i=1 ,i\neq n}^{N}\mathbf{\theta}_{i}\mathbf{b}_{i}\mathbf{a}_{i}^{H}+\mathbf{H}_{ab}\right)\times\notag\\
&~~~~~\left(\sum_{i=1 ,i\neq n}^{N}\mathbf{\theta}_{i}\mathbf{b}_{i}\mathbf{a}_{i}^{H}+\mathbf{H}_{ab}\right)^{H},  \notag\\
&\mathbf{B}_{n}=\rho \mathbf{b}_{n}\mathbf{a}_{n}^{H}\mathbf{a}_{n}\mathbf{b}_{n}^{H}, \notag\\
&\mathbf{C}_{n}=\rho \mathbf{b}_{n}\mathbf{a}_{n}^{H}\left(\mathbf{H}_{ab}^{H}+\sum_{i=1 ,i\neq n}^{N}\mathbf{a}_{i}\mathbf{b}_{i}^{H}\mathbf{\theta}_{i}^{*}\right).  \notag
\end{align}
\end{small}
It can be observed that the matrices $\mathbf{A}_{n}$, $\mathbf{B}_{n}$ and $\mathbf{C}_{n}$ do not contain the variable $\mathbf{\theta}_{n}$, which means that these matrices can be obtained if all the elements $\{\mathbf{\theta}_{i}\}_{i=1,i\neq n}^{N}$ are fixed.

Similarly, the relationship between the transmit power of the relay and $\mathbf{\Psi}$ can be determined as
\begin{align}
P_{r}&=\mathrm{trace}(\mathbf{\Psi}(\mathbf{H}_{rb}\mathbf{H}_{rb}^{H}P_{a}+
                                   \sigma_{b}^{2})\mathbf{\Psi}^{H})\notag\\
                                   &= P_{a}\sum_{n\in \mathbb{Q}}|\mathbf{\psi}_{n}|^{2}||\mathbf{b}_{n}||^{2}+\sigma_{b}^{2}\sum_{n\in \mathbb{Q}}|\mathbf{\psi}_{n}|^{2}\notag\\
                                   &=\sum_{n\in \mathbb{Q}}|\mathbf{\psi}_{n}|^{2}[P_{a}||\mathbf{b}_{n}||^{2}+\sigma_{b}^{2}].  \label{P_r}
\end{align}

Denote $\widetilde{P}_{r}$ = $\sum_{i\in \mathbb{Q},i\neq n}|\mathbf{\psi}_{i}|^{2}[\sigma_{b}^{2}+P_{a}||\mathbf{b}_{n}||^{2}]$, which is a constant due to that the variables $\sum_{i\in \mathbb{Q},i\neq n}\mathbf{\psi}_{i}$ are fixed. Therefore, \eqref{P_r} can be rewritten as
\begin{align}
P_{r}&=\sum_{n\in \mathbb{Q}}|\mathbf{\psi}_{n}|^{2}[P_{a}||\mathbf{b}_{n}||^{2}+\sigma_{b}^{2}]+\widetilde{P}_{r}\notag \\
     &=\sum_{n\in \mathbb{Q}}|\mathbf{\beta}_{n}|^{2}[P_{a}||\mathbf{b}_{n}||^{2}+\sigma_{b}^{2}]+\widetilde{P}_{r}.
\end{align}
Here, we notice that $|\mathbf{\psi}_{n}|^{2}=|\mathbf{\beta}_{n}|^{2}$ for $n\in \mathbb{Q}$.
\subsection{An Efficient Algorithm to Solve $\mathrm{(P2)}$}
\subsubsection{Problem of Updating $\mathbf{\Theta}$} In our proposed algorithm, since $\left\{\theta_{i}\right\}_{i=1,i\notin \mathbb{Q}}^{N}$ is fixed when optimizing $\mathbf{\theta}_{n}$ in each iteration, the objective function $f_{0}({\mathbf{\Theta}},P_{a})$ can be rewritten as
\begin{align}
f_{0}(\mathbf{\Theta},P_{a})&=\log_{2}\Big{|}\mathbf{A}_{n}+|\theta_{n}|^{2}\mathbf{B}_{n}
+\mathbf{\theta}_{n}\mathbf{C}_{n}
+\mathbf{\theta}_{n}^{*}\mathbf{C}_{n}^{H}\Big{|}\notag\\
&=\log_{2}|\mathbf{A}_{n}|+f_{1}(\mathbf{\Theta},P_{a}), \label{f0}
\end{align}
where $\mathbf{A}_{n}$ is an invertible matrix satisfying $\mathrm{rank}(\mathbf{A}_{n})=N_{b}$. Moreover, $\log_{2}(|\mathbf{A}_{n}|)$ is a constant, and $f_{1}(\mathbf{\Theta},P_{a})$ is given by
\begin{small}
\begin{align}
f_{1}({\mathbf{\Theta}},P_{a})=\log_{2}|\mathbf{I}_{N_{b}}+|\theta_{n}|^{2}\mathbf{A}_{n}^{-1}
\mathbf{B}_{n}+\mathbf{\theta}_{n}\mathbf{A}_{n}
^{-1}\mathbf{C}_{n}+\mathbf{\theta}_{n}^{*}\mathbf{A}_{n}^{-1}\mathbf{C}_{n}^{H}|.
\end{align}
\end{small}
Following the above transformation, the problem of updating $\mathbf{\Theta}$, denoted by ($P_{3}$), is given by
\begin{align}
\mathrm{(P3)}:&\mathop{\textup{max}}\limits_{\mathbf{\Theta}} ~f_{1}(\mathbf{\Theta},P_{a})\notag \\
&\mathbf{s.t.} ~|\mathbf{\beta}_{n}|=1,&{\mathrm{for}~n\notin \mathbb{Q},}\notag \\
&~~~~~|\mathbf{\beta}_{n}|^{2}\leq\frac{P_{r}^{\mathrm{max}}-\widetilde{P}_{r}}
{[\sigma_{b}^{2}+P_{a}||\mathbf{b}_{n}||^{2}]},&{\mathrm{for}~n\in \mathbb{Q}.}\label{p3}
\end{align}
\subsubsection{Solution to $\mathrm{(P3)}$}
$\mathrm{(P3)}$ in \eqref{p3} admits a closed-form solution, and, thus, it is efficient for practical implementation. In order to derive it, the objective function $f_{1}(\mathbf{\Theta},P_{a})$ can be rewritten as
\begin{align} \label{f0}
f_{1}(\mathbf{\Theta},P_{a})&=\log_{2}|\mathbf{D}_{n}+\mathbf{\theta}_{n}\mathbf{A}_{n}^{-1}
\mathbf{C}_{n}+\mathbf{\theta}_{n}^{*}\mathbf{A}_{n}^{-1}\mathbf{C}_{n}^{H}| \\
&=\log_{2}|\mathbf{D}_{n}|+\textup{log}_{2}\Big{|}\mathbf{I}_{N_{b}}+\mathbf{\theta}_{n}\mathbf{E}_{n}^{-1}
\mathbf{C}_{n}+\mathbf{\theta}_{n}^{*}\mathbf{E}_{n}^{-1}\mathbf{C}_{n}^{H}\Big{|}, \notag
\end{align}
where $\mathbf{D}_{n}$ = $\mathbf{I}_{N_{b}}+|\theta_{n}|^{2}\mathbf{A}_{n}^{-1}\mathbf{B}_{n}$ and $\mathbf{E}_{n}$ = $\mathbf{A}_{n}\mathbf{D}_{n}$.

We next analyse the objective function $f_{1}(\mathbf{\Theta})$ by considering the first term in \eqref{f0}, i.e., $\log_{2}|\mathbf{D}_{n}|$. Specifically, for $|\mathbf{D}_{n}|$, we note that $\mathrm{rank}(\mathbf{A}_{n}^{-1}\mathbf{B}_{n})$ $<$ $\mathrm{rank}(\mathbf{B}_{n})=1$. Moreover, the probability of $\mathrm{rank}(\mathbf{A}_{n}^{-1}\mathbf{B}_{n})$ is close to zero (it only happens when $\mathbf{A}_{n}^{-1}\mathbf{B}_{n}= 0$). Thus, we have $\mathrm{rank}(\mathbf{A}_{n}^{-1}\mathbf{B}_{n})= 1$. Similarly, we find that $\mathbf{A}_{n}^{-1}\mathbf{B}_{n}$ is not diagonalizable when $\mathrm{rank}(\mathbf{A}_{n}^{-1}\mathbf{B}_{n})= 0$, which usually rarely happens. Based on this, we have $(\mathbf{A}_{n}^{-1}\mathbf{B}_{n})\neq 0$ with a high probability and $\mathbf{A}_{n}^{-1}\mathbf{B}_{n}$ is diagonalizable. Hence, we can rewrite $\mathbf{A}_{n}^{-1}\mathbf{B}_{n}=\mathbf{W}_{n}\mathbf{\Sigma}_{n}\mathbf{W}_{n}^{-1}$ based on EVD, where $\mathbf{\Sigma}_{n}$ = diag$\left\{\iota_{n},0,\cdots,0\right\}$, $\iota_{n}$ is the only non-zero eigenvalue of $(\mathbf{A}_{n}^{-1}\mathbf{B}_{n})$. Finally, since both $\mathbf{A}_{n}$ and $\mathbf{B}_{n}$ are positive semidefinite, $\iota_{n}$ is nonnegative and real. Thus, we have
\begin{align}
\log_{2}|\mathbf{D}_{n}|&=\log_{2}|\mathbf{I}_{N_{b}}+|\theta_{n}|^{2}\mathbf{W}_{n}\mathbf{\Sigma}_{n}
                              \mathbf{W}_{n}^{-1}|\notag\\
                              &=\log_{2}|\mathbf{W}_{n}(\mathbf{I}_{N_{b}}+|\theta_{n}|^{2}\mathbf{\Sigma}_{n})
                              \mathbf{W}_{n}^{-1}|\notag\\
                              &=\log_{2}\Big{|}1+|\theta_{n}|^{2}\iota_{n}\Big{|}.\label{log2Dn}
\end{align}

We are now focusing on the second term of~\eqref{f0}. By a similar argument for the first term, we have $\mathbf{E}_{n}^{-1}\mathbf{C}_{n}$ diagonalizable with a high probability as well. Thus, we have $\mathbf{E}_{n}^{-1}\mathbf{C}_{n}=\mathbf{T}_{n}{\mathbf{\Gamma}}_{n}\mathbf{T}_{n}^{-1}$ based on the EVD, where $\mathbf{T}_{n}\in \mathbb{C}^{N_{b}\times N_{b}}$, $\mathbf{\Gamma}_{n}=\mathrm{diag}\left\{\lambda_{n},0,\cdots,0\right\}$, $\lambda_{n}$ is the sole non-zero eigenvalue of $\mathbf{E}_{n}^{-1}\mathbf{C}_{n}$. Let $\mathbf{V}_{n}=\mathbf{T}_{n}\mathbf{A}_{n}\mathbf{T}_{n}^{-1}$, and $v_{n}$ denote first element of the first column of $\mathbf{V}_{n}^{-1}$ and $v_{n}^{'}$ denote first element of the first row of $\mathbf{V}_{n}$. Note that it follows that $v_{n}^{'}v_{n}=1$. So, according to the \cite{2020CC}, we can write
\begin{align}
\log_{2}|&\mathbf{I}_{N_{b}}+\mathbf{\theta}_{n}\mathbf{E}_{n}^{-1}\mathbf{C}_{n}+\mathbf{\theta}_{n}^{*}
                                   \mathbf{E}_{n}^{-1}\mathbf{C}_{n}^{H}|=\notag\\
                                   &\log_{2}(1+|\theta_{n}|^{2}|\lambda_{n}|^{2}+2\mathbf{\mathbb{R}}
                                   (\mathbf{\theta}_{n}\lambda_{n})
                                   -v_{n}^{'}v_{n}|\lambda_{n}|^{2}),\label{log2INb}
\end{align}where $\mathbb{R}$ denotes the real part of a complex number. We note that the additional coefficient $|\theta_{n}|^{2}$ is related to the active relay elements in HR-RIS, which does not exist in traditional RIS.

In summary, based on~\eqref{log2Dn} and~\eqref{log2INb}, we have
\begin{align}
f_{1}(&\mathbf{\Theta},P_{a})=\log_{2}(1+|\theta_{n}|^{2}\iota_{n})\notag\\
                                   &+\log_{2}(1+|\theta_{n}|^{2}|\lambda_{n}|^{2}+
                                   2\mathbf{\mathbb{R}}(\mathbf{\theta}_{n}
                                   \lambda_{n})-v_{n}^{'}v_{n}|\lambda_{n}|^{2}). \label{f1theta}
\end{align}

Hence, according to \eqref{f1theta} we have $\mu_{n}^{*}=\mathrm{arg}(\lambda_{n})$. So the optimal solution of the problem (P3) is given by
\begin{align}
\mathbf{\theta}_{n}^{*}=
\left\{
  \begin{array}{ll}
     |\beta_{n}|e^{-j\mathrm{arg}(\lambda_{n})}, &{n\in \mathbb{Q},} \\
    e^{-j\mathrm{arg}(\lambda_{n})}, &{n\notin \mathbb{Q}.}\label{betan}
  \end{array}
\right.
\end{align}

In the HR-RIS, $\mathbb{Q}$ is available to determine $\left\{|\beta_{n}|\right\}_{n\in \mathbb{Q}}$. Therefore, from~\eqref{p3}, we obtain
\begin{align}
|\beta_{n}|=\sqrt{\frac{P_{r}^{\mathrm{max}}-\widetilde{P}_{r}}
{[\sigma_{b}^{2}+P_{a}||\mathbf{b}_{n}||^{2}]}},~~n\in \mathbb{Q}.  \label{op_beta}
\end{align}

As a result, the optimal solution to (P3) is given as
\begin{align}
\mathbf{\theta}_{n}^{*}=
\left\{
  \begin{array}{ll}
     \sqrt{\frac{P_{r}^{\mathrm{max}}-\widetilde{P}_{r}}{[\sigma_{b}^{2}+P_{a}
     ||\mathbf{b}_{n}||^{2}]}}e^{-j\mathrm{arg}(\lambda_{n})}, &{n\in \mathbb{Q},} \\
    e^{-j\mathrm{arg}(\lambda_{n})}, &{n\notin \mathbb{Q}.}
  \end{array}\label{sl}
\right.
\end{align}
\begin{remark} \label{remark1}
\textup{Substituting $\widetilde{P}_{r}$=$\sum_{i\in \mathbb{Q},i\neq n}|\mathbf{\beta}_{i}|^{2}[\sigma_{b}^{2}+P_{a}||\mathbf{b}_{n}||^{2}], n\in \mathbb{Q}$ to~\eqref{op_beta}, it is observed that a larger $K$ results in a smaller $|\beta_{n}|$. Therefore, increasing the number of active elements (i.e., $K$) does not always guarantee the covert rate improvement of HR-RIS over traditional RIS. In particular, with a limited power budget $P_{r}^{\mathrm{max}}$, the HR-RIS can have $|\mathbf{\beta}_{n}|< 1$, which attenuates the signal and degrades the covert rate. In this case, HR-RIS with a smaller $K$ is more likely to attain a covert rate gain than those with more active elements. This conclusion will be further demonstrated numerically.} \label{re1}
\end{remark}
\subsection{\small\textup{Optimizing $P_{a}$} for a Given $\mathbf{\Theta}$}
\begin{corollary}\label{corollary1}
KL divergence $\mathcal{D}_{01}$ is a monotonically increasing function of $P_{a}$.
\end{corollary}
\begin{IEEEproof}
In order to determine the monotonicity of $\mathcal{D}_{01}$ with respect to $P_{a}$, we derive its first derivative as
\begin{align} \label{D/Pa}
&\frac{\partial{\mathcal{D}_{01}}}{\partial{P_{a}}}  \\
&=L\left(\frac{P_{a}\mathbf{U}_{w}\mathbf{U}_{w}^{H}}
{\left(\sigma_{w}^{2}\left(\mathbf{I}_{N_{w}}+\mathbf{H}_{rw}\mathbf{\Psi} \mathbf{\Psi}^{H}\mathbf{H}_{rw}^{H}\right)+P_{a}\mathbf{U}_{w}\right)^{2}}\right)>0.  \notag
\end{align}
\end{IEEEproof}
For a given $\mathbf{\Theta}$, following~\eqref{D/Pa}, we can find optimal transmit power of Alice $P_{a}^{*}$ by solving $\mathcal{D}_{01}= 2\epsilon^{2}$.

Since $P_{a}$ does not affect the optimization of $\mathbf{\Theta}$, $P_{a}^{*}$ is the global optimal solution.
\section{Numerical Results and Discussions}
In this section, the numerical results will validate the performance of our proposed scheme. We assume that uniform linear arrays (ULAs) are deployed at the Alice, Bob and Willie, respectively. In contrast, HR-RIS uses a unified plane array (UPA) with $N$ elements. Furthermore, assuming that there is a half-wavelength distance between Alice, Bob, Willie and HR-RIS arrays. We consider a two-dimensional coordinate system, Alice, HR-RIS, Bob, and Willie are respectively located at $(0\mathrm{m},0\mathrm{m})$, $(51\mathrm{m},0\mathrm{m})$, $(50\mathrm{m},2\mathrm{m})$ and $(30\mathrm{m},5\mathrm{m})$. All channel realizations are drawn from Rician fading. The path loss of a link distance $d$ is given by \cite{2019irse,2020CC}, $\chi(d)=\chi_{0}(\frac{d}{1\mathrm{m}})^{\alpha_{ij}}$, where $\chi_{0}$ is the path loss at the reference distance of 1 meter, and $\alpha_{ij}$ is the path loss exponent. Specifically, the path loss exponents are set as $\alpha_{ar}=2.2$, $\alpha_{rb}=2.8$, $\alpha_{ab}=4.2$, $\alpha_{aw}=4.2$, and $\alpha_{rw}=2.8$.

\begin{figure}
  \centering
  \includegraphics[width=3in, height=2.5in]{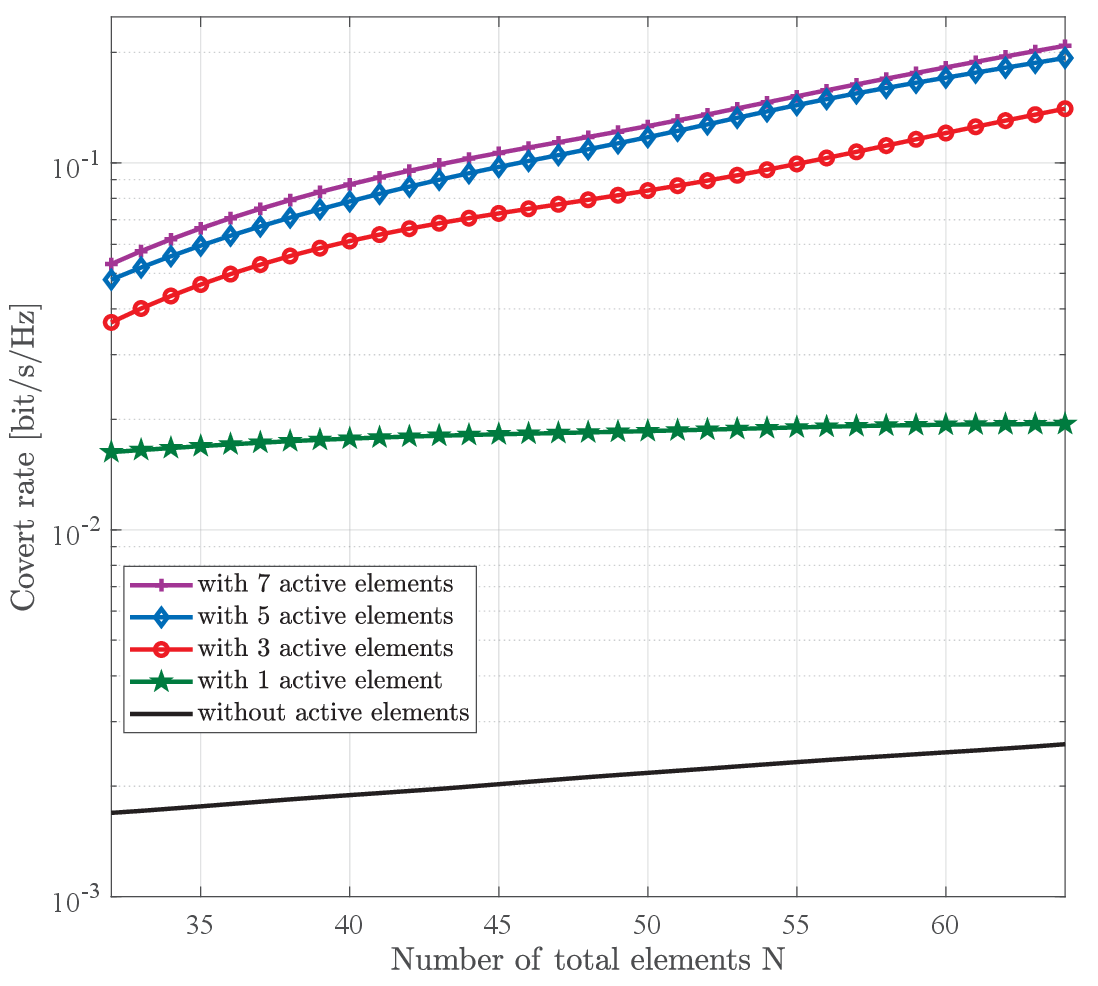}\\
  \caption{Bob's covert rate versus the number of reflecting elements at HR-RIS for different numbers of the active elements $K$,  for $\epsilon=0.01$, $P_{r}^{\mathrm{max}}=-30$~dBm, $L=100$, $\sigma_{b}^{2}=\sigma_{w}^{2}=-80$~dBm.} \label{te}
\end{figure}

In Fig.~\ref{te}, we plot the covert rate of Bob versus the total reflection elements $N$ of the HR-RIS under different numbers of the active elements $K$. In this figure, we first observe that Bob's covert rate increases as $N$ increases. It can also be seen that Bob's covert transmission rate increases as $K$ increases. As the number of active elements increases, we can see that the increase in the covert rate becomes smaller. This is due to the power limit of active elements, too many active elements can cause $|\mathbf{\beta}_{n}|< 1$, which is explained in Remark \ref{remark1}. Therefore, we can get the optimal number of active elements from this figure. Based on which, we can conclude that a small amount of active elements is sufficient for HR-RIS to achieve a significant improvement in term of the covert rate when compared to conventional RIS-aided covert communication schemes.

\section{Conclusion}
In this work, we proposed a covert communication scheme with the help of the HR-RIS, where several elements are active relay elements and the remaining ones as passive reflecting elements. We first used KL divergence to represent Willie's detection capability, and proved that the KL divergence adopted in the covertness constraint is a monotonically increasing function of the transmit power at Alice. Based on this, we obtained the optimal value of $P_a$. Then, we used an alternate optimization algorithm to identify the optimal reflection coefficients of HR-RIS to improve the covert communication performance. The numerical results demonstrated that the proposed scheme can significantly outperform the conventional RIS-aided covert communication schemes in term of covert rate by using a small number of active elements.

\vspace{-0.35cm}
\bibliographystyle{IEEEtran}
\bibliography{IEEEfull,ref}

\begin{thebibliography}{10}
\providecommand{\url}[1]{#1}
\csname url@samestyle\endcsname
\providecommand{\newblock}{\relax}
\providecommand{\bibinfo}[2]{#2}
\providecommand{\BIBentrySTDinterwordspacing}{\spaceskip=0pt\relax}
\providecommand{\BIBentryALTinterwordstretchfactor}{4}
\providecommand{\BIBentryALTinterwordspacing}{\spaceskip=\fontdimen2\font plus
\BIBentryALTinterwordstretchfactor\fontdimen3\font minus
  \fontdimen4\font\relax}
\providecommand{\BIBforeignlanguage}[2]{{%
\expandafter\ifx\csname l@#1\endcsname\relax
\typeout{** WARNING: IEEEtran.bst: No hyphenation pattern has been}%
\typeout{** loaded for the language `#1'. Using the pattern for}%
\typeout{** the default language instead.}%
\else
\language=\csname l@#1\endcsname
\fi
#2}}
\providecommand{\BIBdecl}{\relax}
\BIBdecl

\bibitem{2019Low}
S.~Yan, X.~Zhou, J.~Hu, and S.~V. Hanly, ``Low probability of detection
  communication: Opportunities and challenges,'' \emph{IEEE Wireless Commun.},
  vol.~26, no.~5, pp. 19--25, Oct. 2019.

\bibitem{Hu2018covertrelay}
J.~Hu, S.~Yan, X.~Zhou, F.~Shu, J.~Li, and J.~Wang, ``Covert communication
  achieved by a greedy relay in wireless networks,'' \emph{IEEE Trans. Wireless
  Commun.}, vol.~17, no.~7, pp. 4766--4779, Jul. 2018.

\bibitem{2020Intelligent}
X.~Lu, E.~Hossain, T.~Shafique, S.~Feng, H.~Jiang, and D.~Niyato, ``Intelligent
  reflecting surface enabled covert communications in wireless networks,''
  \emph{IEEE Netw.}, vol.~34, no.~5, pp. 148--155, Oct. 2020.

\bibitem{2020bo}
Q.~Wu and R.~Zhang, ``Beamforming optimization for wireless network aided by
  intelligent reflecting surface with discrete phase shifts,'' \emph{IEEETrans.
  Commun.}, vol.~68, no.~3, pp. 1838--1851, Mar. 2020.

\bibitem{2021MIMO}
X.~Chen, T.-X. Zheng, L.~Dong, M.~Lin, and J.~Yuan, ``Enhancing {MIMO} covert
  communications via intelligent reflecting surface,'' \emph{IEEE Wireless
  Commun. Lett.}, vol.~11, no.~1, pp. 33 -- 37, Jan. 2022.

\bibitem{2020DC}
X.~Zhou, S.~Yan, Q.~Wu, F.~Shu, and D.~W.~K. Ng, ``Intelligent reflecting
  surface {IRS}-aided covert wireless communication with delay constraint,''
  \emph{IEEE Trans. Wireless Commun.}, vol.~21, no.~1, pp. 532--547, Jan. 2022.

\bibitem{2019Intelligent}
E.~Bjornson, O.~Ozdogan, and E.~Larsson, ``Intelligent reflecting surface
  versus decode-and-forward: How large surfaces are needed to beat relaying?''
  \emph{IEEE Wireless Commun. Lett.}, vol.~9, no.~2, pp. 244 -- 248, Feb. 2020.

\bibitem{2021Hybrid}
N.~T. Nguyen, Q.~D. Vu, K.~Lee, and M.~Juntti, ``Hybrid relay-reflecting
  intelligent surface-assisted wireless communication,'' \emph{IEEE Trans. Veh.
  Technol.}, Mar. 2022, {E}arly Access.

\bibitem{2019irse}
Q.~Wu and R.~Zhang, ``Intelligent reflecting surface enhanced wireless network
  via joint active and passive beamforming,'' \emph{IEEE Trans. Wireless
  Commun.}, vol.~18, no.~11, pp. 5394 -- 5409, Nov. 2019.

\bibitem{2020RISce}
Z.~Wang, L.~Liu, and S.~Cui, ``Channel estimation for intelligent reflecting
  surface assisted multiuser communications: Framework, algorithms, and
  analysis,'' \emph{IEEE Trans. Wireless Commun.}, vol.~19, no.~10, pp.
  6607--6620, Oct. 2020.

\bibitem{2018Delay}
S.~Yan, B.~He, X.~Zhou, Y.~Cong, and A.~L. Swindlehurst, ``Delay-intolerant
  covert communications with either fixed or random transmit power,''
  \emph{IEEE Trans. Inf. Forensics Security.}, vol.~14, no.~1, pp. 129 -- 140,
  Jan. 2019.

\bibitem{2020CC}
S.~Zhang and R.~Zhang, ``Capacity characterization for intelligent reflecting
  surface aided {MIMO} communication,'' \emph{IEEE J. Sel. Areas Commun.},
  vol.~38, no.~8, pp. 1823--1838, Aug. 2020.

\end{thebibliography}

\end{document}